\documentclass[sigconf]{acmart}
\AtBeginDocument{%
  \providecommand\BibTeX{{%
    \normalfont B\kern-0.5em{\scshape i\kern-0.25em b}\kern-0.8em\TeX}}}

\setcopyright{acmcopyright}

\copyrightyear{2024}
\acmYear{2024}
\setcopyright{acmlicensed}\acmConference[WSDM '24]{Proceedings of the 17th ACM International Conference on Web Search and Data Mining}{March 4--8, 2024}{Merida, Mexico}
\acmBooktitle{Proceedings of the 17th ACM International Conference on Web Search and Data Mining (WSDM '24), March 4--8, 2024, Merida, Mexico}
\acmPrice{15.00}
\acmDOI{10.1145/3616855.3635778}
\acmISBN{979-8-4007-0371-3/24/03}

\usepackage{ulem}
\usepackage{subcaption}
\usepackage{ulem}
\usepackage{multirow}
\usepackage{graphicx}
\usepackage{algorithm, algpseudocode}

\begin{document}

\title{Budgeted Embedding Table For Recommender Systems}
\author{Yunke Qu}
\affiliation{
  \institution{The University of Queensland}
  \city{Brisbane}
  \state{}
  \country{Australia}
}
\email{yunke.qu@uq.net.au}

\author{Tong Chen}
\affiliation{
  \institution{The University of Queensland}
  \city{Brisbane}
  \state{}
  \country{Australia}
}
\email{tong.chen@uq.edu.au}

\author{Quoc Viet Hung Nguyen}
\affiliation{
  \institution{Griffith University}
  \city{Gold Coast}
  \state{}
  \country{Australia}
}
\email{henry.nguyen@griffith.edu.au}

\author{Hongzhi Yin}
\authornote{Corresponding author}
\affiliation{
  \institution{The University of Queensland}
  \city{Brisbane}
  \state{}
  \country{Australia}
}
\email{h.yin1@uq.edu.au}

\renewcommand{\shortauthors}{Yunke Qu, Tong Chen, Quoc Viet Hung Nguyen, \& Hongzhi Yin}

\begin{abstract}
At the heart of contemporary recommender systems (RSs) are latent factor models that provide quality recommendation experience to users. These models use embedding vectors, which are typically of a uniform and fixed size, to represent users and items. As the number of users and items continues to grow, this design becomes inefficient and hard to scale. Recent lightweight embedding methods have enabled different users and items to have diverse embedding sizes, but are commonly subject to two major drawbacks. Firstly, they limit the embedding size search to optimizing a heuristic balancing the recommendation quality and the memory complexity, where the trade-off coefficient needs to be manually tuned for every memory budget requested. The implicitly enforced memory complexity term can even fail to cap the parameter usage, making the resultant embedding table fail to meet the memory budget strictly. Secondly, most solutions, especially reinforcement learning based ones derive and optimize the embedding size for each each user/item on an instance-by-instance basis, which impedes the search efficiency. In this paper, we propose Budgeted Embedding Table (BET), a novel method that generates table-level actions (i.e., embedding sizes for all users and items) that is guaranteed to meet pre-specified memory budgets. Furthermore, by leveraging a set-based action formulation and engaging set representation learning, we present an innovative action search strategy powered by an action fitness predictor that efficiently evaluates each table-level action. Experiments have shown state-of-the-art performance on two real-world datasets when BET is paired with three popular recommender models under different memory budgets. Code is available at https://github.com/qykcq/Budgeted-Embedding-Table-For-Recommender-Systems.
\end{abstract}



\maketitle

\vspace{-0.5em}
\section{Introduction}
Recommender systems (RSs) predict a user's preference for an item according to observed user-item interactions \cite{zhang2019deep, wang2021survey} and have been indispensable across numerous modern platforms such as e-commerce and social media \cite{rececom1, rececom2, kywe2012survey}. In contexts where computations on local devices are essential, such as federated recommendation \cite{muhammad2020fedfast} and Internet of Things (IoT) services \cite{huang2019multimodal}, the need for RSs with small memory footprints becomes critical due to highly constrained computational resources. The main bottleneck for downsizing RSs is often the embedding table, which represents users and items with distinct embedding vectors to facilitate pairwise similarity computation. As such, the size of the embedding table can quickly exhaust the available memory, posing a significant challenge to achieving scalable and resource-friendly RSs. 

Such memory bottleneck is attributed to conventional embedding tables, which typically allocate fixed and uniform embedding sizes to each user and item \cite{joglekar2020neural, chentongkdd2021}. Such a design may require an excessive number of parameters in industry-scale applications, e.g., the benchmark recommender in \cite{DesaiS22} has more than 25 billion parameters in its embedding table and requires 100 GB memory. In this regard, researchers have developed methods to achieve lightweight embeddings while maintaining its expressiveness. One stream of methods aim to prune redundant parameters of the embedding table. For example, \cite{optemb} proposed OptEmbed, a novel embedding sparsification framework that prunes redundant embeddings by learning an importance score for each feature. These methods mostly optimize a loss function combining a recommendation objective term and a sparsity regularizer. Apart from sparsification/pruning methods, another line of works featuring reinforcement learning (RL) techniques have also emerged. Those methods adopt an RL-based policy network to search for the optimal embedding sizes for each user/item, where the embedding sizes (i.e., actions) can be selected from an action space consisting of either a predefined set of values \cite{esapn} or a continuous range \cite{ciess}. Similar to sparsification methods, the reward functions in RL-based lightweight embedding approaches also bind the recommendation quality term with a memory complexity penalty. By tuning a trade-off coefficient among the two terms, they can adjust the regularization strength on the parameter consumption, thus achieving the desired memory efficiency.

Despite the possibility of balancing performance and embedding parameter size, these two types of methods are still away from practical use for two reasons. Firstly, such a desired balance is achieved by jointly optimizing an additional space complexity constraint on top of the recommendation objective. However, such an implicit loss term means that the resultant embedding table may not necessarily meet the desired parameter budget every time \cite{chentongkdd2021}. Moreover, the optimization quality is highly sensitive to the trade-off coefficient, which needs to be heuristically tuned for each memory budget.

The second reason is that, to achieve optimal recommendation performance under each memory budget, the embedding size is decided on an instance-by-instance basis w.r.t. each user/item. 
Taking a recent RL-based method CIESS \cite{ciess} as an example, a policy network predicts the optimal action for every single user and item, based on which the backbone RS is trained and evaluated to calculate the reward, enabling training the policy network. On the one hand, ideally, the policy network is expected to be iteratively updated w.r.t. each instance (i.e., user/item), but this can be prohibitively inefficient given the large number of users and items. On the other hand, performing only one update per batch/epoch leads to slower convergence due to the aggregated nature of the reward function, and may potentially result in inferior generalization \cite{keskar2016large}.

\begin{figure}
    \centering
    \includegraphics[width=\linewidth]{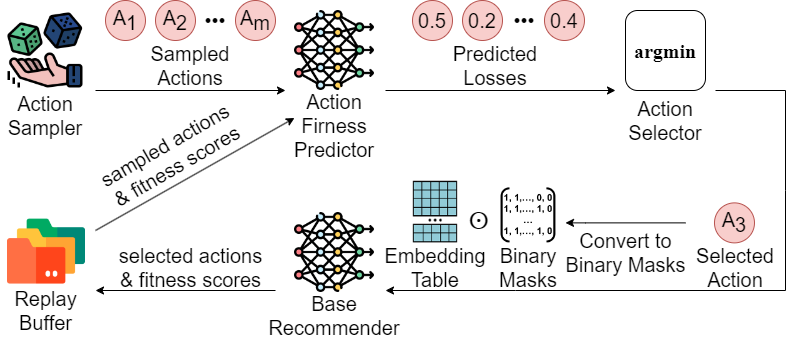}
    \vspace{-1em}
    \caption{Overview of BET.} \label{fig:overview}
    \vspace{-1.5em}
\end{figure}

In response to both challenges, we propose Budgeted Embedding Table (BET) for RSs, an automated input embedding size search paradigm that can efficiently generate size decisions and perform optimization at the embedding table level. An overview of BET is provided in Figure \ref{fig:overview}. To address the first challenge, we innovatively propose a bounded sampler that draws embedding sizes for all users/items from a collection of probabilistic distributions, with the guarantee that the total parameter usage is capped. As the embedding sizes for all instances in the embedding table are produced all at once, we term this a \textit{table-level action}. This strategy frees us from the need for manually designing and tuning the implicit parameter constraint in the reward/loss function, and allows us to enforce the memory budget with a negligible cost. 

To tackle the second challenge, we further pair the generated table-level actions with a fitness prediction network to evaluate the quality of those actions. Essentially, in each search iteration, after sampling several table-level actions, we need to identify the ones that promise high recommendation accuracy. A straightforward solution is to greedily evaluate each candidate action by training and evaluating a backbone RS from scratch with the suggested embedding sizes \cite{ciess, zhaok2021autoemb, joglekar2020neural}, which apparently impedes efficiency. 

One may think of a surrogate function \cite{chentongkdd2021, cai2020once} that ``predicts'' the fitness of each action, instead of explicitly evaluating it. Accurately predicting the performance of the table-level actions is crucial yet another major challenge in this regard, as either overestimating or underestimating an action's performance will eventually cause suboptimal embedding size decisions. To align with the evaluation of table-level actions in BET, our fitness predictor innovatively utilizes a set-based formulation of actions, and embeds each action with DeepSets \cite{deepsets}, where the learned high-quality action embeddings can efficiently provide reliable performance feedback on each table-level action. Furthermore, the set-based action representations can be inductively learned, which is essential for our iterative search algorithm that constantly generates unseen actions that need to be evaluated. 

To summarize, our work presents the contributions below:
\begin{itemize}
    \item We identify the practicality bottlenecks of existing embedding size search methods for lightweight RSs, namely the implicit and non-scalable parameter size constraints for meeting different memory budgets, and the inefficient embedding size generation and evaluation for every instance.
    \item We propose BET, a novel table-level embedding size search approach featuring an efficient action sampling strategy that generates table-level actions for all users and items and a DeepSets-based fitness prediction network that accurately predicts their performance to reduce action evaluation costs.
    \item Extensive empirical comparisons have been carried out on two real-world datasets by pairing state-of-the-art embedding size search algorithms with various backbone RSs, which confirm the advantageous efficacy of BET.
\end{itemize}

\section{Preliminaries}
To provide a detailed research background, we hereby discuss relevant research topics.

\subsection{Deep Recommender Systems}
Researchers have proposed numerous deep recommender models to capture the user-item relationships. The first line of deep recommender systems are based on MLPs. For instance, Neural Collaborative Filtering (NCF) \cite{he2017neural} integrates a MLP with a matrix factorization component to learn the two-way interactions between users and items. AutoRec \cite{autorec} uses an autoencoder to model the user-item interactions in a low-dimensional hidden space and perform collaborative filtering. Lin et al. proposed a novel sequential recommender system, Dual Contrastive Network \cite{Lin2022DualCN} that generates self-supervised signals by considering both user and item perspectives and refining representation learning through dual representation contrastive learning. In addition to MLP-based methods, graph-based architectures \cite{wang2019neural, he2020lightgcn, graphaug} have also been studied. For example, Neural Graph Collaborative Filtering (NGCF) \cite{wang2019neural} extends traditional neural collaborative filtering by modeling user-item interactions with graph convolution networks. LightGCN \cite{he2020lightgcn} takes inspiration from NGCF and simplifies it by removing its self-connection, feature transformation and linear activation operations. Another line of recommender systems are based on factorization machines. Factorization Machines (FM) combine linear regression with matrix factorization techniques to handle high-order interactions \cite{rendle2010factorization}. \cite{hofm} was also created to mine higher-order interactions. Its deep variants - QNFM \cite{qnfm}, DeepFM \cite{guo2017deepfm} and xLightFM \cite{xlightfm} have been proposed too.
Moreover, numerous methods have been proposed to leverage heterogeneous data for recommendation. For instance, \cite{trythis, gong2016hashtag, li2017neural} have been devised to extract information from user reviews or hashtags. \cite{lee2018collaborative, suhangwang2017www} have been proposed to exploit videos and images for recommendation. Location information has also been utilized to assist location-aware recommendations \cite{chen2017iteeaccess}. 

\subsection{Lightweight Embeddings For Recommenders}
 Various classes of works have investigated learning compressed embeddings to provide a low-memory alternative to traditional embedding tables in recommender systems \cite{automlrecsurvey}. For examples, feature selection search methods \cite{wang2022autofield, luo2019autocross, adafs} are trained to learn importance scores, based on which they filter out the unimportant feature fields. Embedding dimension search approaches \cite{liu2021learnable, esapn, zhaok2021autoemb, singleshot, autosrh, ciess} improve traditional recommender systems with fixed and uniform sizes by introducing mixed dimensions for each feature. Quantization-based methods \cite{dpq, mascot, kang2020www, li2023aaai, chen2022learning, chen2022kdd} reduces the precision of numerical values and store the parameters with lower bits. Feature hashing approaches \cite{yan2021cikm, zhang2020model, sethi2022asplos, liu2022tkde, shi2020compositional, li2021lightweight, Desai22mlsys, DesaiS22, icdm23xurong} utilize various hash tricks to transform input values to a smaller range, thus compression the embedding table. Knowledge distillation techniques train a smaller student model from a larger teacher model and have been used to learn an embedding table of smaller sizes from larger ones \cite{xiaxin2022, xiaxin2023}. BET achieves dynamic embedding sizes and aims to find the optimal embedding size for each feature, so it should be classified as a embedding dimension search approach. 

\section{Proposed Approach}
In a nutshell, BET has three components: (1) a pre-trained backbone recommender with a full-size embedding table $G_{\Theta}(\cdot)$ parameterized by $\Theta$; (2) an action fitness prediction network $F_{\Phi}(\cdot)$ parameterized by $\Phi$; and (3) a non-parametric action sampler $H(\cdot)$. Figure \ref{fig:overview} provides an overview of the workflow of BET. Under pre-specified memory budgets, $H(\cdot)$ samples actions from a set of probabilistic distributions in each iteration. Each action specifies the embedding sizes for all users and items. 
Afterwards, $F_{\Phi}(\cdot)$ estimates the fitness of each action and selects the best action predicted. $G_{\Theta}(\cdot)$ is further finetuned by adopting the embedding sizes in the chosen action, and is then evaluated on a hold-out dataset to obtain the actual fitness score (i.e., recommendation quality) of the action. This facilitates the optimization of $F_{\Phi}(\cdot)$ by minimizing its prediction error, thus leading us to optimal actions under each given memory budget after $T$ iterations. In the remainder of this section, we unfold the design of each component.

\subsection{Recommender with Embedding Masking} \label{subsec:baserec}
Let $\mathcal{U}$ and $\mathcal{V}$ denote a set of users $u$ and items $v$, respectively. The embedding table in the recommender maps each integer-valued user or item IDs to a real-valued embedding vector $\mathbf{e}_{n}$, where $n\in\mathcal{U}\cup \mathcal{V}$ indexes either a user or an item. Note that we only separately use $u$ and $v$ to index users and items when they need to be distinguished, e.g., when performing pairwise ranking. The vertical concatenation of all embedding vectors forms the embedding table $\mathbf{E}$ of size $(|\mathcal{U}|+|\mathcal{V}|) \times d_{max}$, where $d_{max}$ is the maximum/full embedding size supported by $G_{\Theta}(\cdot)$. The embedding sizes are adjusted via a binary mask $\mathbf{M}\in \{0,1\}^{(|\mathcal{U}|+|\mathcal{V}|) \times d_{max}}$, which is commonly adopted for embedding sparsification in lightweight recommender systems \cite{liu2021learnable, optemb, ciess}. Each row $\mathbf{m}_n$ in $\mathbf{M}$ is a mask vector corresponding to user/item $n$. Given a searched embedding size $d_n$, we set the first $d_n$ elements of $\mathbf{m}_n$ to ones the remaining $d_{max} - d_n$ dimensions to zeros. Then, we perform element-wise multiplication of the embedding table $\mathbf{E}$ and the binary mask matrix $\mathbf{M}$ to adjust the embedding sizes during embedding look-up, i.e., $\mathbf{e}_{n} = (\mathbf{E} \odot \mathbf{M})[n]$ where $[n]$ retrieves the $n$-th row of a matrix.  $\mathbf{M}$ is adaptively updated based on the sampled actions in every iteration.
It is worth noting that, in the deployment stage, only the heavily sparsified embedding table $\mathbf{E}_{sparse} = \mathbf{E} \odot \mathbf{M}$ will be stored, where the cost of storing zero-valued entries can be negligible with sparse matrix storage techniques \cite{virtanen2020scipy, sedaghati2015automatic}. Once we obtain the sparse embedding vectors, the recommender model $G_{\Phi}(\cdot)$ computes a preference score $\hat{y}_{uv}$ representing pairwise user-item similarity:
\begin{equation}
    \hat{y}_{uv} = G_{\Theta}(\mathbf{e}_u,\mathbf{e}_{v}), \;\;\;\; u\in\mathcal{U}, v\in \mathcal{V},
\end{equation}
where $G_{\Theta}(\cdot)$ can be any latent factor-based recommender model that supports pairwise scoring.  

For optimizing the recommender, we adopt Bayesian Personalized Ranking (BPR) loss \cite{rendle_bpr_2012}:
\begin{equation} \label{eq:bpr}
   \mathcal{L}_{BPR} = \sum_{(u, v, v') \in \mathcal{D}_{train}} - \ln \sigma (\hat{y}_{uv} - \hat{y}_{uv'}) + \eta\|\Theta\|_2^2,
\end{equation}
where $(u, v, v') \in \mathcal{D}_{train}$ represents the training samples where user $u$ chooses item $v$ over item $v'$, while $\hat{y}_{uv}$ and $\hat{y}_{uv'}$ respectively represent the predicted preferences that user $u$ has on items $v$ and $v'$. 
$\eta\|\Theta\|_2^2$ provides regularization to prevent overfitting, with a scaling factor $\eta$. Our goal is to search for the optimal embedding sizes for all users and items under strict memory budgets, hence the overall objective is defined as follows:
\begin{equation}\label{eq:objective}
    \min_{\Theta,\Phi} \mathcal{L}_{BPR} \;\;\;\text{s.t.}\;1- \frac{\| \mathbf{M}\|_{1,1}}{(|\mathcal{U}|+|\mathcal{V}|) \times d_{max}} \leq c,
\end{equation}
where $\| \mathbf{M}\|_{1,1}$ counts the number of ones, or equivalently retained parameters in the binary mask, $c$ ($0< c <1$) is the target sparsity rate, i.e., the ratio between the numbers of active parameters in sparsified and full embedding tables.

\subsection{Budget-Aware Action Sampling}
Hereby we introduce a novel table-level action sampling strategy. In each search iteration, we randomly sample a set of candidate actions $\mathcal{Q} = \{a_1, ..., a_m\}$. Each action assigns a continuous embedding size $d_{n} \in \mathbb{N}^+_{\leq d_{max}}$ to every user or item $n \in\mathcal{U}\cup \mathcal{V}$. 

As we aim to avoid optimizing the selected action towards an implicit, manually tuned memory cost term, we propose to directly draw a table-level action that: (1) makes all generated embedding sizes conditioned on a certain distribution; and (2) ensures the total parameter size of all embeddings is capped at the budgeted parameter consumption. In short, (1) essentially narrows the action space to mitigate the difficulty of searching from a huge collection of completely random actions, while (2) guarantees that the table-level action strictly meets the memory budget in the first place.

To generate an action $a$, we first determine two probability distributions $P_{\mathcal{U}}$ and $P_{\mathcal{V}}$ respectively for users and items. 
Meanwhile, to introduce inductive bias to the action sampler, instead of resorting to a fixed distribution throughout the search process, $P_{\mathcal{U}}$ and $P_{\mathcal{V}}$ are both randomly chosen from a set of candidate distributions with on-the-fly parameterization. Specifically, we utilize a power law distribution, a truncated exponential distribution, a truncated normal distribution and a log normal distribution. For simplicity, we let the two selected distributions respectively be parameterized by $\beta_{\mathcal{U}}$ and $\beta_{\mathcal{V}}$.
In each iteration, the parameters are all determined uniformly at random:
\begin{equation}\label{eq:uvdistribution}
\beta_{\mathcal{U}}, \beta_{\mathcal{V}} \sim \text{Uniform}(0, \beta_{max}),
\end{equation}
where $\beta_{max}$ is the maximal value allowed. Notably, the embedding sizes of each field (i.e., users and items) are derived by independently sampling from two distributions $P_{\mathcal{U}}$ and $P_{\mathcal{V}}$. This is because each field may have distinct properties \cite{optemb} and different distributions are needed to model the embedding sizes of users and items. 

After the distributions are determined, we sample $|\mathcal{U}|$ and $|\mathcal{V}|$ probabilities respectively from $P_{\mathcal{U}}$ and $P_{\mathcal{V}}$. The set of probabilities are respectively denoted as  $\{p_i\}_{i=1}^{|\mathcal{U}|}\sim P_{\mathcal{U}}$ and $\{p_j\}_{j=1}^{|\mathcal{V}|}\sim P_{\mathcal{V}}$. 
We then normalize all $p_i$ and $p_j$ as the following:
\begin{equation}\label{eq:pi_pj}
 \tilde{p}_i = \frac{p_i}{\sum_i^{|\mathcal{U}|} p_i}, \,\,\,\,
 \tilde{p}_j = \frac{p_j}{\sum_j^{|\mathcal{V}|} p_j},
\end{equation}
where $\tilde{p}_i$/$\tilde{p}_j$ indicates the fraction of parameters each user/item receives among all users/items. Next, we obtain the actual embedding sizes by applying the fraction on the memory budget:
\begin{equation}\label{eq:actions}
\begin{split}
 d_i = \lfloor \tilde{p}_i \cdot w \cdot d_{max} \cdot (|\mathcal{U}| + |\mathcal{V}|) \cdot c \rfloor, \;\;\; i = 1,2,\cdots,|\mathcal{U}|,\\
 d_j = \lfloor \tilde{p}_j \cdot (1 - w) \cdot d_{max} \cdot (|\mathcal{U}| + |\mathcal{V}|) \cdot c \rfloor, \;\;\; j = 1,2,\cdots,|\mathcal{V}|,
\end{split}
\end{equation}
where $w$ controls the proportion of parameters allocated to the users. In practice, $w$ can be empirically defined, such as an even split (i.e., 0.5). In BET, $w$ is sampled from a uniform distribution each time a table-level action is generated, i.e., $w \sim \text{Uniform}(0, 1)$, so as to probe more embedding size allocation strategies in the iterative search process. 

With $|\mathcal{U}| + |\mathcal{V}|$ embedding sizes derived, we now allocate them to each user and item. 
Previous studies \cite{zhaok2021autoemb, optemb, cprec} have shown that more frequent users and items tend to be more informative than long-tail ones, hence generally benefit from a larger embedding dimension which allows more information to be encoded. 
Therefore, we rank the users by their frequency and allocate the $n$-th largest user embedding size to the $n$-th frequent user. Similarly, we assign embedding sizes to items based on their frequency. As such, we are capable of sampling $m$ random table-level actions in each search iteration to facilitate the subsequent steps. 

\vspace{-0.2cm}

\subsection{Fitness Prediction Network}

In each iteration $t$ we sample a set of $m$ actions, denoted as $\mathcal{Q}_t$. To select the optimal action from $\mathcal{Q}_t$, a straightforward approach is to pair each sampled action with a fresh recommender, train it until convergence and select the action that yields the best final recommendation quality on the validation set. However, this approach faces inevitable computational burdens given the $m$ training and evaluation processes needed per search iteration. 

In BET, we utilize an action fitness predictor, which inductively learns distinct embeddings of actions and predicts its fitness score (i.e., recommendation quality) based on the action embeddings. As discussed earlier, learning expressive representations of different actions is a crucial yet non-trivial task for effective fitness prediction, especially considering the need for inductively composing embeddings for unseen actions in every new search iteration. 

\begin{figure} 
    \centering
    \includegraphics[width=0.4\linewidth]{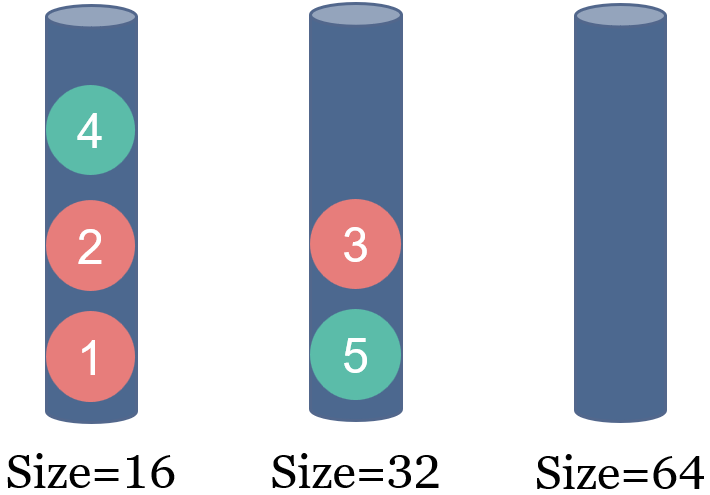}
    \vspace{-1em}
    \caption{Example of our set-based action formulation. Suppose we have three legitimate embedding sizes $\{16, 32, 64\}$, four users (red) and two items (green). Then, size 16 is a set containing user 1, user 2 and item 4, which means their embedding sizes are 16. Similarly, the action specifies the embedding sizes of user 3 and item 5 to be 32. Size 64 is not used by any users/items.} \label{fig:action}
    \vspace{-1em}
\end{figure}

\begin{figure}
    \centering
    \includegraphics[width=0.8\linewidth]{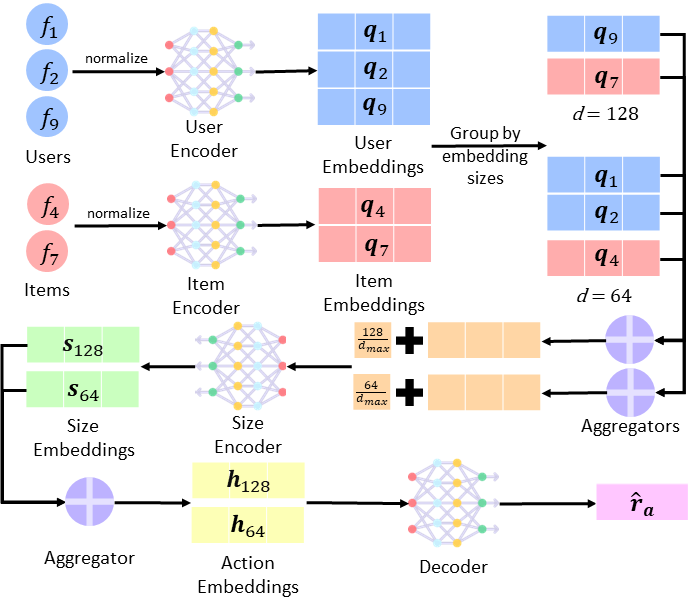}
    \vspace{-1.0em}
    \caption{Overview of the DeepSets-based fitness predictor.} \label{fig:deepsets}
    \vspace{-1.3em}
\end{figure}

Thus, we propose a novel action fitness prediction network based on the notion of set representation learning \cite{deepsets, wagstaff2022universal}. Before delving into the fitness prediction network, we first detail the set-based formulation of each action. Each action is regarded as a collection of sets $a=\{\mathcal{S}_d\}_{d=1}^{d_{max}}$, where each set $\mathcal{S}_d=\{n|\forall n\in \mathcal{U}\cup\mathcal{V}, d_n = d\}$ stores the users and items that use the embedding size $d\in [1, d_{max}]$. $\mathcal{S}_d = \emptyset$ if there is no user or item assigned to embedding size $d$. A toy example with is given in figure \ref{fig:action}. As such, this allows us to build our fitness predictor upon DeepSets \cite{deepsets} owing to its ability to learn permutation-variant representations of sets. 

The fitness prediction network consists of four components: a user encoder $\rho_{\mathcal{U}}(\cdot)$, an item encoder $\rho_{\mathcal{V}}(\cdot)$, an embedding size encoder $\mu(\cdot)$, and a decoder $\pi(\cdot)$, all of which bear a distinct MLP. Encoders $\rho_{\mathcal{U}}(\cdot)$/$\rho_{\mathcal{V}}(\cdot)$ take some contextual information about a user/item, and produce their corresponding representations. 
Previous studies \cite{zhaok2021autoemb, ciess} have demonstrated the suitability of users'/items' popularity and current embedding sizes as the context information for embedding size search. 
We inherit this design by establishing a hierarchical, set-compatible action representation learning scheme. Specifically,   $\rho_{\mathcal{U}}(\cdot)$ and $\rho_{\mathcal{V}}(\cdot)$ respectively take the user and item frequencies to generate frequency-weighted embeddings:
\begin{equation}
\begin{split}
    \boldsymbol{q}_{u} &= \rho_{\mathcal{U}}(\frac{f_{u}}{\max_{u'\in\mathcal{U}}f_{u'}}), \;\;\; u \in \mathcal{U}, \\ 
    \boldsymbol{q}_{v} &= \rho_{\mathcal{V}}(\frac{f_{v}}{\max_{v'\in\mathcal{V}}f_{v'}}), \;\;\; v \in \mathcal{V},
\end{split}
\end{equation}
where each $f_u$/$f_v$ is normalized by the maximum user/item frequency observed. 

As mentioned above, within an action $a$, each set $\mathcal{S}_d$ contains all users and items with embedding size $d$. To obtain embedding $\boldsymbol{s}_d$ of set $\mathcal{S}_d$, we firstly fuse the representations of all involved users and items using mean aggregation, and then append the explicit embedding size information:
\begin{equation}
    \boldsymbol{s}_d = \mu\Big{(}\Big{[}\frac{1}{|\mathcal{S}_d|}\sum_{n \in \mathcal{S}_d} \boldsymbol{q}_n; \frac{d}{d_{max}}\Big{]}\Big{)},
\end{equation}
where $[;]$ represents concatenation. With the representations generated for all $d_{max}$ sets, we derive the embedding for the entire action $a$ by further merging all set representations:
\begin{equation}\label{eq:action_emb}
    \boldsymbol{h}_a = \frac{1}{d_{max}}\sum_{d=1}^{d_{max}} \boldsymbol{s}_d, 
\end{equation}
where we use $\boldsymbol{h}_a$ to denote an action embedding. An additional benefit from the set-based action embeddings is that, whenever the action sampler draws any action that is new to the fitness predictor (which is commonly the case in embedding size search), it allows for generating the action embedding in an inductive fashion with the available context information and the updated set formulation.

With each action embedding $\boldsymbol{h}_a$, the decoder $\pi(\cdot)$ takes $\boldsymbol{h}_a$ and outputs $\hat{r}_a$, which is the predicted fitness score of action $a$:
\begin{equation} \label{eq:fitness prediction}
    \hat{r}_a = \pi(\boldsymbol{h}_a).
\end{equation}
We will detail how the entire fitness prediction network $F_{\Phi}(\cdot)$ is optimized along with the search process in the next section. 

\subsection{Fitness Predictor Training}
Though bearing a relatively simple structure, the fitness predictor still needs to be trained with some ground truth samples. The fitness prediction network $F_{\Phi}(\cdot)$ is optimized by minimizing the mean squared error between the predicted fitness score $\hat{r}_a$ and the actual one $r_a$ with gradient descent:
\begin{equation} \label{eq:mse}
    \Phi = \underset{\Phi'}{\text{argmin}} \sum_{\forall a}(r_a - F_{\Phi'}(a))^2 = \underset{\Phi'}{\text{argmin}} \sum_{\forall a}(r_a - \hat{r}_a)^2.
\end{equation}

The actual fitness score $r_a$, i.e., recommendation quality of the action, can be evaluated on the backbone recommender $G_{\Theta}(\cdot)$ after being finetuned with an action $a$ w.r.t. Eq. \ref{eq:bpr}:
\begin{equation} \label{eq:fitness}
    r_a = \frac{\text{eval}(\mathbf{E} \odot \mathbf{M}_a|\mathcal{D}_{val})}{\text{eval}(\mathbf{E}|\mathcal{D}_{val})},
\end{equation}
where $\text{eval}(\cdot)$ evaluates the recommendation quality, $\mathcal{D}_{val}$ is the validation set, $\mathbf{M}_a$ is the binary mask corresponding to action $a$, and the denominator is the recommendation quality of the pre-trained backbone recommender $G_\Theta(\cdot)$ with the full embedding table. Hence, $r_a$ can be interpreted as the ratio between the two recommendation accuracy obtained by the sparsified embeddings and the full embeddings. The implementation of $\text{eval}(\cdot)$ can follow most commonly used recommendation metrics like Recall@$k$ and NDCG@$k$ \cite{he2020lightgcn,rendle_bpr_2012}, where we adopt an ensemble of both metrics under different values of $k$:
\begin{equation}\label{eq:eval}
    \text{eval}(\mathbf{E} \odot \mathbf{M}|\mathcal{D}_{val}) = \sum_{u \in \mathcal{U}} \frac{\sum_{k \in \mathcal{K}} \text{Recall@}k_u + \text{NDCG@}k_u}{2|\mathcal{K}||\mathcal{U}|},
\end{equation}
where we the choices of $k$ cover $\mathcal{K}=\{5,10,20\}$ in our paper. 

The ground truth samples can be thus generated by gathering different action-fitness pairs. Instead of constructing a pool of such training samples and train $F_{\Phi}(\cdot)$ upfront, we incrementally add one training sample in every search iteration $t$ to optimize $F_{\Phi}(\cdot)$ on the go. The training sample is constructed by firstly identifying a promising action from $\mathcal{Q}_t$, denoted by $a_t$, and then obtaining $r_{a_t}$ by evaluating $a_t$ on the finetuned $G_{\Theta}(\cdot)$. Compared with most RL-based embedding size search methods, the required number of fitness evaluation with the recommendation model will be reduced from $m \times T$ to merely $T$, where $T$ is the total number of iterations as well as the ground truth samples for optimizing $F_{\Phi}(\cdot)$. 

\begin{algorithm}[t]
\caption{Embedding Size Search with BET}
\label{alg:bet}
\begin{algorithmic}[1]
\State Initialize $\Theta$ and $\Phi$, set $\mathcal{A}\leftarrow \emptyset$;
\State Train $G_\Theta(\cdot)$ till convergence w.r.t. Eq. \ref{eq:bpr};
\For{$t = 1,\cdots,T$}
    \State Draw distributions and their parameters w.r.t. Eq. \ref{eq:uvdistribution};
    \State Draw $m$ actions $\mathcal{Q}_t = \{a_1, ..., a_m\}$ with Eq. \ref{eq:pi_pj} and Eq. \ref{eq:actions};
    \If{$0 \leq t \text{ mod } 5 \leq 2$}
        \State $a_{t} \leftarrow$ Strategy I;
    \ElsIf{$t \text{ mod } 5 = 3$}
        \State $a_{t} \leftarrow$ Strategy II;
    \ElsIf{$t \text{ mod } 5 = 4$}
        \State $a_{t} \leftarrow$ Strategy III;
    \EndIf    
    \State Finetune $\Theta$ w.r.t. $\textbf{E}\odot \textbf{M}_{a_t}$ and Eq. \ref{eq:bpr}
    \State $r_{a_t} \leftarrow$ Eq. \ref{eq:eval}
    \State $\mathcal{A} \leftarrow \mathcal{A} \cup (a_{t}, r_{a_t})$;
    \For{$j =1,\cdots,N$}
        \State Randomly sample $(a_j,r_{a_j})$ from $\mathcal{A}$;
        \State $\hat{r}_{a_j} \leftarrow $ Eq. \ref{eq:fitness prediction};
        \State Update $\Phi$ w.r.t. Eq. \ref{eq:mse};
    \EndFor
\EndFor
\State Perform selective retraining and obtain $a^* \leftarrow$ Eq. \ref{eq:selective}.
\end{algorithmic}
\end{algorithm}

\subsection{Action Selection and Embedding Retraining}
The pseudocode for BET is presented in Algorithm \ref{alg:bet}. Among its $T$ search iterations, we utilize three different strategies to pick one action $a_t$ from the candidate action set $\mathcal{Q}_t$ (lines 4-12). Then, the recommender $G_{\Theta}(\cdot)$ is finetuned by adopting $a_t$ and evaluated on the validation set (lines 13-14), where the produced $(a_t, r_{a_t})$ sample is popped into the population set $\mathcal{A}$ to facilitate training the fitness prediction network $F_{\Phi}(\cdot)$ (lines 16-20). Finally, a selective retraining process is conducted with the top-ranked actions in the population in order to identify the best embedding size decision for all users and items. We provide further details about action selection strategies and retraining the sparsified embeddings below. 

\subsubsection{Diversifying the Action Space.}  The iterative input embedding size search contains $T$ iterations. In each iteration $t$, the selected action $a_t$ is added to the population set $\mathcal{A}$ with its associated fitness score. For efficiency consideration, in each iteration $t$, we only compute the actual fitness score for action $a_t$, which is selected from candidate actions $\mathcal{Q}_t$ with one of the three following strategies:
\begin{enumerate}
    \item[I:] We use our fitness prediction network $F_\Phi(\cdot)$ to predict the fitness score of each candidate actions, and greedily selects the action predicted to have the highest fitness score, i.e., $a_{t} = \text{argmax}_{a \in \mathcal{Q}_t} F_{\Phi}(a)$. In essence, $F_\Phi(\cdot)$ is a surrogate function that is trained to approximate Eq. \ref{eq:fitness} and is substantially more cost-effective than evaluating all $m$ actions via recommender finetuning. 
    \item[II:] We randomly select one action from the $\mathcal{Q}_t$ to diversify the training samples for $F_\Phi(\cdot)$ by preventing it from overfitting some specific action distributions in the population $\mathcal{A}$.
    \item[III:] We identify currently the best action (with the highest $r_a$) in $\mathcal{A}$. All actions in $\mathcal{Q}_t$ and the best action are inductively embedded into vector representations using Eq. \ref{eq:action_emb}. Then, by calculating the pairwise Euclidean distance, we select action $a_t\in \mathcal{Q}_t$ by locating the best action's nearest neighbor in the embedding space. 
\end{enumerate}

\subsubsection{Selective Retraining for Sparsified Embeddings.} \label{sec:retraining} After $T$ search iterations, from the population we identify the top five actions with the highest $r_a$ on the validation set, denoted as $\mathcal{A}'$. Each action $a\in \mathcal{A}'$ is paired with a fresh backbone recommender model, which is trained from scratch on the training set until convergence. Next, we evaluate the trained recommenders and select the final action $a^*$ that yields the best validation performance:
\begin{equation} \label{eq:selective}
    a^* = \underset{a \in \mathcal{A}'}{\text{argmax}}\,\, r_a,
\end{equation}
with which the sparsified embedding $\textbf{E}_{sparse}$ is obtained and stored for inference.  

\section{Experiments}
This section details the experimental results of BET's performance.

\subsection{Base Recommenders and Comparative Methods}
One can combine BET with different embedding-based representation learning backbone recommender models. We tested the effectiveness of our approach with three commonly used recommender models as the backbone recommender $F_{\Phi}$: NCF \cite{he2017neural}, NGCF \cite{wang2019neural} and LightGCN \cite{he2020lightgcn}. We used the same settings as reported in their original works and replaced the embedding table with one with adjustable embedding sizes to demonstrate the versatility and generalizability of our method across various backbone recommenders. We compare BET with the following model-agnostic embedding size search algorithms:
\begin{itemize}
    \item PEP \cite{liu2021learnable}: It adaptively prunes the embedding parameters based on a learned pruning threshold learned along with the model's other parameters during the training process.
    \item ESAPN \cite{esapn}: It uses an RL agent which performs hard selection on embedding sizes for users and items based on the memory complexity and recommendation performance.
    \item OptEmbed \cite{optemb}: It prunes redundant embeddings based on learnable pruning thresholds indicative of features' importance and derives the optimal embedding size for each field using evolutionary search.
    \item CIESS \cite{ciess}: It is an RL-based method that chooses embedding sizes for the users and items from a continuous domain.
    \item Equal Sizes (ES): All users and items share equal and fixed embedding sizes.
    \item Mixed and Random (MR): The embedding sizes of the users and items are sampled from a uniform distribution.
\end{itemize}

\subsection{Evaluation Protocols}
All the aforementioned methods are evaluated on two real-world datasets: Gowalla \cite{gowalla} and Yelp2018 \cite{wang2019neural}. The first has 1,027,370 interactions between 29,858 users and 40,981 items while the latter has 1,561,147 interactions between 31,668 users and 38,048 items. We take 50\%, 25\% and 25\% of the two datasets for training, validation, and test. To measure the recommendation performance of the backbone recommender, Recall@$k$ and NDCG@$k$ are used as evaluation metrics, where $k$ is set to either 5 or 20. To test the recommendation performance of CIESS, PEP, ES and MR, three different sparsity ratios ($c$) of 80\%, 90\% and 95\% are used. For each method, the compressed embedding table is ensured to have no more than $c \cdot d_{max} \cdot(|\mathcal{U}|+|\mathcal{V}|)$ usable parameter and the maximal embedding size is $d_{max}=128$. However, since ESAPN and OptEmbed are not designed to precisely control the embedding sparsity and focus more on the performance, this section only reports the performance associated with their final embedding tables after pruning.

\begin{table*}[th]
\renewcommand*{\arraystretch}{0.88}
\resizebox{\textwidth}{!}{%
\begin{tabular}{cccccccccccccccc}
\hline
 & \multicolumn{5}{c|}{LightGCN} & \multicolumn{5}{c|}{NGCF} & \multicolumn{5}{c}{NCF} \\ \hline
\multicolumn{1}{c|}{Method} & Sparsity & R@5 & N@5 & R@20 & \multicolumn{1}{c|}{N@20} & Sparsity & R@5 & N@5 & R@20 & \multicolumn{1}{c|}{N@20} & Sparsity & R@5 & N@5 & R@20 & N@20 \\ \hline
\multicolumn{1}{c|}{ESAPN} & 75\% & 0.0741 & 0.1173 & 0.1579 & \multicolumn{1}{c|}{0.1329} & 83\% & 0.0627 & 0.1060 & 0.1396 & \multicolumn{1}{c|}{0.1193} & 74\% & 0.0498 & 0.0718 & 0.1280 & 0.0959 \\ \hline
\multicolumn{1}{c|}{OptEmb} & 74\% & 0.0591 & 0.0983 & 0.1314 & \multicolumn{1}{c|}{0.1115} & 75\% & 0.0510 & 0.0860 & 0.1227 & \multicolumn{1}{c|}{0.1005} & 82\% & 0.0330 & 0.0492 & 0.0877 & 0.0653 \\ \hline
\multicolumn{1}{c|}{PEP} & \multirow{5}{*}{80\%} & 0.0715 & 0.1189 & 0.1588 & \multicolumn{1}{c|}{0.1347} & \multirow{5}{*}{80\%} & 0.0656 & 0.1049 & \textbf{0.1524} & \multicolumn{1}{c|}{0.1245} & \multirow{5}{*}{80\%} & 0.0498 & 0.0704 & 0.1282 & 0.0957 \\
\multicolumn{1}{c|}{ES} &  & 0.0756 & 0.1219 & 0.1674 & \multicolumn{1}{c|}{0.1400} &  & 0.0619 & 0.0961 & 0.1466 & \multicolumn{1}{c|}{0.1170} &  & 0.0481 & 0.0662 & 0.1235 & 0.0916 \\
\multicolumn{1}{c|}{MR} &  & 0.0708 & 0.1150 & 0.1583 & \multicolumn{1}{c|}{0.1323} &  & 0.0512 & 0.0815 & 0.1223 & \multicolumn{1}{c|}{0.0982} &  & 0.0478 & 0.0656 & 0.1243 & 0.0910 \\
\multicolumn{1}{c|}{CIESS} &  & 0.0777 & 0.1263 & 0.1750 & \multicolumn{1}{c|}{0.1461} &  & \textbf{0.0676} & \textbf{0.1084} & 0.1504 & \multicolumn{1}{c|}{\textbf{0.1251}} &  & \textbf{0.0550} & \textbf{0.0744} & \textbf{0.1307} & \textbf{0.0997} \\
\multicolumn{1}{c|}{BET} &  & \textbf{0.0797} & \textbf{0.1303} & \textbf{0.1753} & \multicolumn{1}{c|}{\textbf{0.1483}} &  & 0.0643 & 0.1039 & 0.1454 & \multicolumn{1}{c|}{0.1204} &  & 0.0511 & 0.0723 & 0.1294 & 0.0973 \\ \hline
\multicolumn{2}{c}{$p$-value} & 0.0024 & 0.0001 & 0.0391 & 0.0012 &  & 0.7815 & 0.2179 & 0.0514 & 0.0924 &  & 0.1962 & 0.1337 & 0.2578 & 0.0414 \\ \hline
\multicolumn{1}{c|}{PEP} & \multirow{5}{*}{90\%} & 0.0594 & 0.0992 & 0.1274 & \multicolumn{1}{c|}{0.1102} & \multirow{5}{*}{90\%} & 0.0552 & 0.0914 & 0.1327 & \multicolumn{1}{c|}{0.1081} & \multirow{5}{*}{90\%} & 0.0414 & 0.0631 & 0.1121 & 0.0839 \\
\multicolumn{1}{c|}{ES} &  & 0.0646 & 0.1041 & 0.1477 & \multicolumn{1}{c|}{0.1217} &  & 0.0589 & 0.0935 & 0.1380 & \multicolumn{1}{c|}{0.1115} &  & 0.0413 & 0.0548 & 0.1112 & 0.0796 \\
\multicolumn{1}{c|}{MR} &  & 0.0617 & 0.1013 & 0.1403 & \multicolumn{1}{c|}{0.1171} &  & 0.0466 & 0.0764 & 0.1123 & \multicolumn{1}{c|}{0.0911} &  & 0.0399 & 0.0536 & 0.1069 & 0.0769 \\
\multicolumn{1}{c|}{CIESS} &  & 0.0721 & 0.1193 & 0.1589 & \multicolumn{1}{c|}{0.1353} &  & 0.0601 & 0.0980 & 0.1380 & \multicolumn{1}{c|}{0.1140} &  & \textbf{0.0448} & 0.0613 & \textbf{0.1131} & 0.0841 \\
\multicolumn{1}{c|}{BET} &  & \textbf{0.0736} & \textbf{0.1212} & \textbf{0.1620} & \multicolumn{1}{c|}{\textbf{0.1376}} &  & \textbf{0.0623} & \textbf{0.1012} & \textbf{0.1426} & \multicolumn{1}{c|}{\textbf{0.1177}} &  & 0.0434 & \textbf{0.0648} & 0.1122 & \textbf{0.0845} \\ \hline
\multicolumn{2}{c}{$p$-value} & 0.0165 & 0.0481 & 0.0091 & 0.0226 &  & 0.0005 & 0.0205 & 0.0601 & 0.0232 &  & 0.1454 & 0.0554 & 0.6608 & 0.0655 \\ \hline
\multicolumn{1}{c|}{PEP} & \multirow{5}{*}{95\%} & 0.0372 & 0.0666 & 0.0803 & \multicolumn{1}{c|}{0.0712} & \multirow{5}{*}{95\%} & 0.0247 & 0.0471 & 0.0674 & \multicolumn{1}{c|}{0.0554} & \multirow{5}{*}{95\%} & 0.0351 & 0.0591 & 0.0897 & 0.0712 \\
\multicolumn{1}{c|}{ES} &  & 0.0446 & 0.0727 & 0.1085 & \multicolumn{1}{c|}{0.0872} &  & 0.0527 & 0.0870 & \textbf{0.1213} & \multicolumn{1}{c|}{\textbf{0.1007}} &  & 0.0336 & 0.0478 & 0.0952 & 0.0682 \\
\multicolumn{1}{c|}{MR} &  & 0.0500 & 0.0834 & 0.1178 & \multicolumn{1}{c|}{0.0975} &  & 0.0438 & 0.0739 & 0.1040 & \multicolumn{1}{c|}{0.0859} &  & 0.0329 & 0.0489 & 0.0918 & 0.0674 \\
\multicolumn{1}{c|}{CIESS} &  & 0.0513 & 0.0853 & 0.1214 & \multicolumn{1}{c|}{0.0997} &  & 0.0505 & 0.0856 & 0.1129 & \multicolumn{1}{c|}{0.0964} &  & 0.0361 & 0.0567 & 0.0933 & 0.0717 \\
\multicolumn{1}{c|}{BET} &  & \textbf{0.0627} & \textbf{0.1037} & \textbf{0.1374} & \multicolumn{1}{c|}{\textbf{0.1171}} &  & \textbf{0.0528} & \textbf{0.0883} & 0.1172 & \multicolumn{1}{c|}{0.9980} &  & \textbf{0.0388} & \textbf{0.0616} & \textbf{0.0963} & \textbf{0.0756} \\ \hline
\multicolumn{2}{c}{$p$-value} & 0.0001 & 0.0002 & 0.0001 & 0.0001 &  & 0.1896 & 0.0638 & 0.7269 & 0.3608 &  & 0.0321 & 0.1107 & 0.1104 & 0.0010 \\ \hline
\multicolumn{16}{c}{(a) Results on Gowalla} \\ \hline
 & \multicolumn{5}{c|}{LightGCN} & \multicolumn{5}{c|}{NGCF} & \multicolumn{5}{c}{NCF} \\ \hline
\multicolumn{1}{c|}{Method} & Sparsity & R@5 & N@5 & R@20 & \multicolumn{1}{c|}{N@20} & Sparsity & R@5 & N@5 & R@20 & \multicolumn{1}{c|}{N@20} & Sparsity & R@5 & N@5 & R@20 & N@20 \\ \hline
\multicolumn{1}{c|}{ESAPN} & 73\% & 0.0294 & 0.0696 & 0.0812 & \multicolumn{1}{c|}{0.0773} & 71\% & 0.0214 & 0.0538 & 0.0632 & \multicolumn{1}{c|}{0.0603} & 76\% & 0.0153 & 0.0342 & 0.0490 & 0.0430 \\ \hline
\multicolumn{1}{c|}{OptEmb} & 80\% & 0.0183 & 0.0424 & 0.0534 & \multicolumn{1}{c|}{0.0489} & 67\% & 0.0133 & 0.0301 & 0.0412 & \multicolumn{1}{c|}{0.0369} & 77\% & 0.0076 & 0.0161 & 0.0238 & 0.0203 \\ \hline
\multicolumn{1}{c|}{PEP} & \multirow{4}{*}{80\%} & 0.0253 & 0.0605 & 0.0723 & \multicolumn{1}{c|}{0.0682} & \multirow{4}{*}{80\%} & 0.0086 & 0.0167 & 0.0275 & \multicolumn{1}{c|}{0.0223} & \multirow{4}{*}{80\%} & 0.0130 & 0.0236 & 0.0406 & 0.0327 \\
\multicolumn{1}{c|}{ES} &  & 0.0289 & 0.0665 & 0.0822 & \multicolumn{1}{c|}{0.0758} &  & 0.0223 & 0.0517 & 0.0665 & \multicolumn{1}{c|}{0.0603} &  & 0.0131 & 0.0226 & 0.0465 & 0.0351 \\
\multicolumn{1}{c|}{MR} &  & 0.0253 & 0.0588 & 0.0783 & \multicolumn{1}{c|}{0.0698} &  & 0.0212 & 0.0499 & 0.0618 & \multicolumn{1}{c|}{0.0573} &  & 0.0101 & 0.0183 & 0.0342 & 0.0268 \\
\multicolumn{1}{c|}{CIESS} &  & 0.0292 & 0.0692 & 0.0839 & \multicolumn{1}{c|}{0.0783} &  & 0.0233 & 0.0566 & 0.0701 & \multicolumn{1}{c|}{0.0652} &  & 0.0175 & 0.0377 & 0.0533 & 0.0474 \\
\multicolumn{1}{c|}{BET} &  & \textbf{0.0310} & \textbf{0.0726} & \textbf{0.0873} & \multicolumn{1}{c|}{\textbf{0.0816}} &  & \textbf{0.0254} & \textbf{0.0601} & \textbf{0.0741} & \multicolumn{1}{c|}{\textbf{0.0685}} &  & \textbf{0.0194} & \textbf{0.0451} & \textbf{0.0620} & \textbf{0.0554} \\ \hline
\multicolumn{2}{c}{$p$-value} & 0.0008 & 0.0309 & 0.0004 & 0.0109 &  & 0.0412 & 0.0418 & 0.0082 & 0.0024 &  & 0.0387 & 0.0122 & 0.0483 & 0.0341 \\ \hline
\multicolumn{1}{c|}{PEP} & \multirow{5}{*}{90\%} & 0.0224 & 0.0531 & 0.0657 & \multicolumn{1}{c|}{0.0610} & \multirow{5}{*}{90\%} & 0.0080 & 0.0153 & 0.0276 & \multicolumn{1}{c|}{0.0215} & \multirow{5}{*}{90\%} & 0.0139 & 0.0242 & 0.0427 & 0.0335 \\
\multicolumn{1}{c|}{ES} &  & 0.0230 & 0.0544 & 0.0722 & \multicolumn{1}{c|}{0.0648} &  & 0.0205 & 0.0528 & 0.0622 & \multicolumn{1}{c|}{0.0592} &  & 0.0119 & 0.0201 & 0.0416 & 0.0309 \\
\multicolumn{1}{c|}{MR} &  & 0.0210 & 0.0476 & 0.0642 & \multicolumn{1}{c|}{0.0573} &  & 0.0195 & 0.0449 & 0.0573 & \multicolumn{1}{c|}{0.0524} &  & 0.0089 & 0.0170 & 0.0309 & 0.0242 \\
\multicolumn{1}{c|}{CIESS} &  & 0.0263 & 0.0649 & 0.0730 & \multicolumn{1}{c|}{0.0705} &  & 0.0232 & 0.0551 & 0.0669 & \multicolumn{1}{c|}{0.0619} &  & 0.0153 & 0.0350 & 0.0500 & 0.0442 \\
\multicolumn{1}{c|}{BET} &  & \textbf{0.0284} & \textbf{0.0674} & \textbf{0.0816} & \multicolumn{1}{c|}{\textbf{0.0762}} &  & \textbf{0.0246} & \textbf{0.0586} & \textbf{0.0736} & \multicolumn{1}{c|}{\textbf{0.0676}} &  & \textbf{0.0173} & \textbf{0.0400} & \textbf{0.0565} & \textbf{0.0499} \\ \hline
\multicolumn{2}{c}{$p$-value} & 0.0045 & 0.0060 & 0.0051 & 0.0006 &  & 0.0012 & 0.0007 & 0.0102 & 0.0069 &  & 0.0234 & 0.0067 & 0.0426 & 0.0118 \\ \hline
\multicolumn{1}{c|}{PEP} & \multirow{5}{*}{95\%} & 0.0199 & 0.0496 & 0.0600 & \multicolumn{1}{c|}{0.0561} & \multirow{5}{*}{95\%} & 0.0075 & 0.0143 & 0.0259 & \multicolumn{1}{c|}{0.0206} & \multirow{5}{*}{95\%} & 0.0125 & 0.0244 & 0.0399 & 0.0330 \\
\multicolumn{1}{c|}{ES} &  & 0.0217 & 0.0494 & 0.0624 & \multicolumn{1}{c|}{0.0571} &  & 0.0196 & 0.0496 & 0.0573 & \multicolumn{1}{c|}{0.0545} &  & 0.0091 & 0.0174 & 0.0303 & 0.0238 \\
\multicolumn{1}{c|}{MR} &  & 0.0195 & 0.0465 & 0.0583 & \multicolumn{1}{c|}{0.0540} &  & 0.0175 & 0.0415 & 0.0528 & \multicolumn{1}{c|}{0.0486} &  & 0.0078 & 0.0145 & 0.0275 & 0.0216 \\
\multicolumn{1}{c|}{CIESS} &  & 0.0230 & 0.0534 & 0.6570 & \multicolumn{1}{c|}{0.0640} &  & \textbf{0.0222} & \textbf{0.0540} & \textbf{0.0662} & \multicolumn{1}{c|}{\textbf{0.0613}} &  & 0.0157 & 0.0383 & 0.0483 & 0.0446 \\
\multicolumn{1}{c|}{BET} &  & \textbf{0.0255} & \textbf{0.0596} & \textbf{0.0727} & \multicolumn{1}{c|}{\textbf{0.0676}} &  & 0.0216 & 0.0505 & 0.0639 & \multicolumn{1}{c|}{0.0588} &  & \textbf{0.0170} & \textbf{0.0404} & \textbf{0.0545} & \textbf{0.0491} \\ \hline
\multicolumn{2}{c}{$p$-value} & 0.0403 & 0.0200 & 0.0305 & 0.0204 &  & 0.0327 & 0.0451 & 0.0064 & 0.0193 &  & 0.0399 & 0.0246 & 0.0094 & 0.0044 \\ \hline
\multicolumn{16}{c}{(b) Results on Yelp} \\
\end{tabular}%
}
\caption{Performance of all methods on (a) Gowalla, and (b) Yelp2018 with the sparsity ratio set to 80\%, 90\%, and 95\%. The best results are highlighted. Recall@$k$ and NDCG@$k$ are denoted by R@$k$ and N@$k$ for simplicity. Each $p$-value is calculated via the paired $t$-test between BET and the best baseline.}
\label{tab:overall}
\vspace{-2.5em}
\end{table*}

\begin{table}[th]
\renewcommand*{\arraystretch}{0.9}
\resizebox{\columnwidth}{!}{%
\begin{tabular}{cccccc}
\hline
\multirow{2}{*}{Sparsity} & \multirow{2}{*}{Selection Method} & \multicolumn{2}{c}{Yelp2018} & \multicolumn{2}{c}{Gowalla} \\ \cline{3-6} 
 &  & R@20 & N@20 & R@20 & N@20 \\ \hline
\multirow{3}{*}{80\%} & Random Selection & 0.0846 & 0.0788 & 0.1749 & 0.1473 \\
 & Simple fitness prediction & 0.0835 & 0.0779 & 0.1739 & 0.1469 \\
 & DeepSets fitness prediction & \textbf{0.0873} & \textbf{0.0816} & \textbf{0.1753} & \textbf{0.1483} \\ \hline
\multirow{3}{*}{90\%} & Random Selection & 0.0767 & 0.0717 & 0.1594 & 0.1357 \\
 & Simple fitness prediction & 0.0768 & 0.0717 & 0.1590 & 0.1355 \\
 & DeepSets fitness prediction & \textbf{0.0816} & \textbf{0.0762} & \textbf{0.1620} & \textbf{0.1376} \\ \hline
\multirow{3}{*}{95\%} & Random Selection & 0.0637 & 0.0583 & 0.1335 & 0.1143 \\
 & Simple fitness prediction & 0.0673 & 0.0629 & 0.1349 & 0.1152 \\
 & DeepSets fitness prediction & \textbf{0.0727} & \textbf{0.0676} & \textbf{0.1374} & \textbf{0.1171} \\ \hline
\end{tabular}%
}
\caption{Performance of BET when different selection methods are used.}
\label{tab:abalation}
\vspace{-3em}
\end{table}

\subsection{Implementation Notes for BET}
Hereby we detail the implementation of BET. The number of search iterations $T$ is 50. The maximal values of the distribution parameters $\beta_{max}$ are set to 20 for power law distribution, 20 for truncated normal distribution, 5 for truncated exponential distribution and 0.5 for log normal distribution. The maximal embedding size $d_{max}$ is 128 and the minimal embedding size $d_{min}$ is 1. In each iteration, the pre-trained backbone recommender is finetuned for 10 epochs (i.e., $h = 10$) when the backbone recommender is LightGCN or NCF. When NGCF is used, the backbone recommender is finetuned for 15 epochs (i.e., $h = 15$). For finetuning, the initial learning rate is 0.03 and decays every 200 steps with a decay ratio of 0.98. The number of fitness prediction updates in each iteration, or $N$, is set to 2. The user and item encoders have input size 1, output size 16, and hidden size 16. The size encoder has input size 17, output size 64, and hidden size 64. The decoder has input size 64, output size 1, and hidden size 64. All the MLP-based encoders and decoders have two layers with the Leaky ReLU \cite{maas2013rectifier} activation function. 

\subsection{Overall Performance Comparison}

Table \ref{tab:overall} shows the Recall@$k$ and NDCG@$k$ scores of various lightweight embedding methods when they are paired with different backbone recommender models. We also perform paired student's $t$-test between BET and the best baseline method. Compared with the baselines with adjustable sparsity ratios (i.e., PEP, ES, MR, and CIESS), BET achieves better performance across all three backbone recommenders and at nearly all specified sparsity levels (80\%, 90\%, and 95\%). The associated $p$-values are below 5\% most of the time so we can reject the null hypothesis that the observed performance changes occurred by chance in most scenarios.

Meanwhile, the sparsity ratios achieved by ESAPN and OptEmbed range from 67\% and 83\%, often failing to meet the minimum target of 80\%. Despite keeping a relatively higher number of parameters compared to other methods, their recommendation performance is consistently inferior to BET. For instance, when used with LightGCN on the Gowalla dataset, ESAPN requires more memory complexity (75\% sparsity ratio) to achieve a comparable performance compared to BET under an 90\% sparsity condition.

In summary, BET demonstrates superior performance while maintaining the same memory usage. Similarly, when comparing performance levels, BET proves to be more efficient in memory utilization. Consequently, these findings highlight the benefits of the ongoing embedding size exploration in BET.

\subsection{Model Component Analysis}

Selecting from the set of candidate actions in each iteration is a vital step in BET, where we proposed a hybrid selection method that alternately uses three selection strategies. To study the effectiveness of the proposed selection method and the DeepSets-based fitness prediction network, we conduct a quantitative study on BET and its two variants. The first variant uses random selection only. The second variant incorporates a simplistic fitness prediction network that only accepts the embedding sizes of each user and item. Table \ref{tab:abalation} illustrates the performance of the original version of BET along with its two variants in terms of their Recall@20 and NDCG@20 scores. Only the performance of the best-performing backbone recommender LightGCN is reported due to space limit. The results show that BET with a DeepSets-based fitness prediction network has better performance than the two variants, implying that our proposed selection method is effective and the DeepSets-based architecture is able to learn the latent representation of actions. 

\subsection{Analysis of Hyperparameter}

In this section, we inspect the influence of the core hyperparameters in BET in terms of Recall@20 and NDCG@20. LightGCN shows the best performance so it is chosen as the backbone recommender for demonstration.

\subsubsection{Number of Sampled Actions $m$}

\begin{figure}[]
    \centering
    \begin{subfigure}{0.92\linewidth} 
        \centering
        \includegraphics[width=\linewidth]{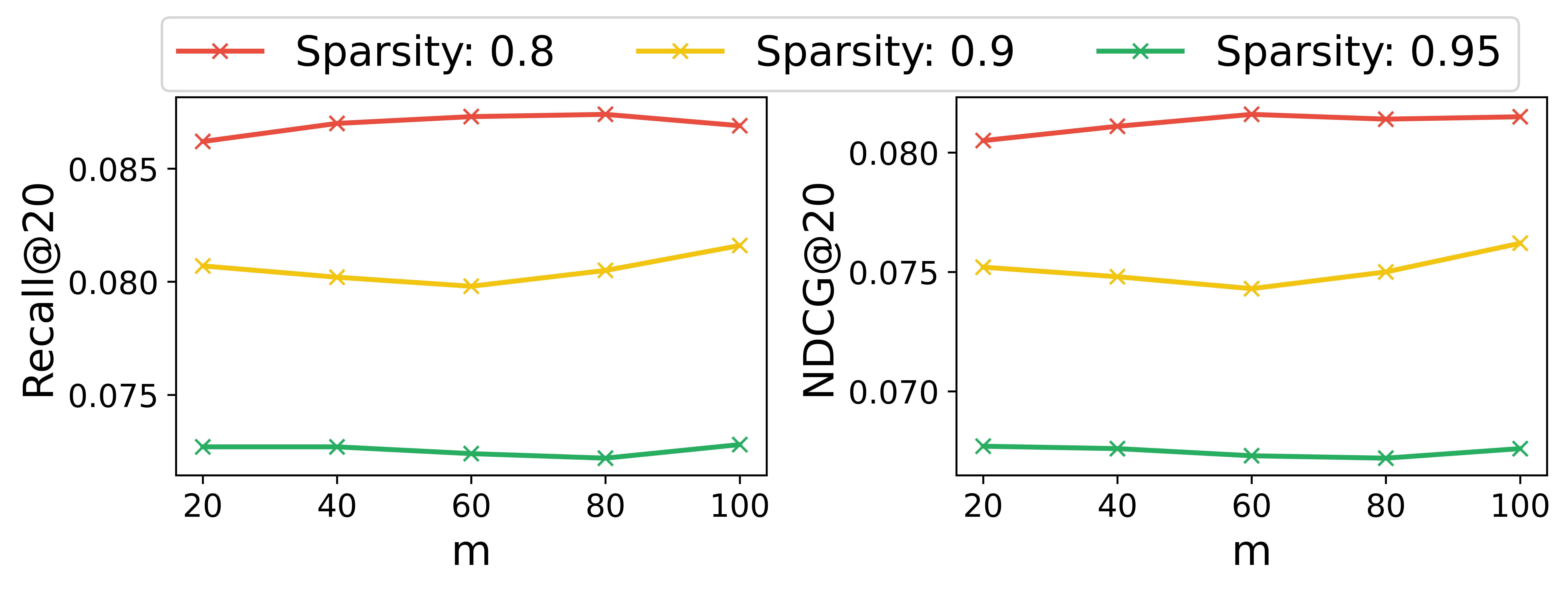}
        \vspace{-2em}
        \caption{Effect of $m$ on Yelp}
    \end{subfigure}
    \begin{subfigure}{0.92\linewidth} 
        \centering
        \includegraphics[width=\linewidth]{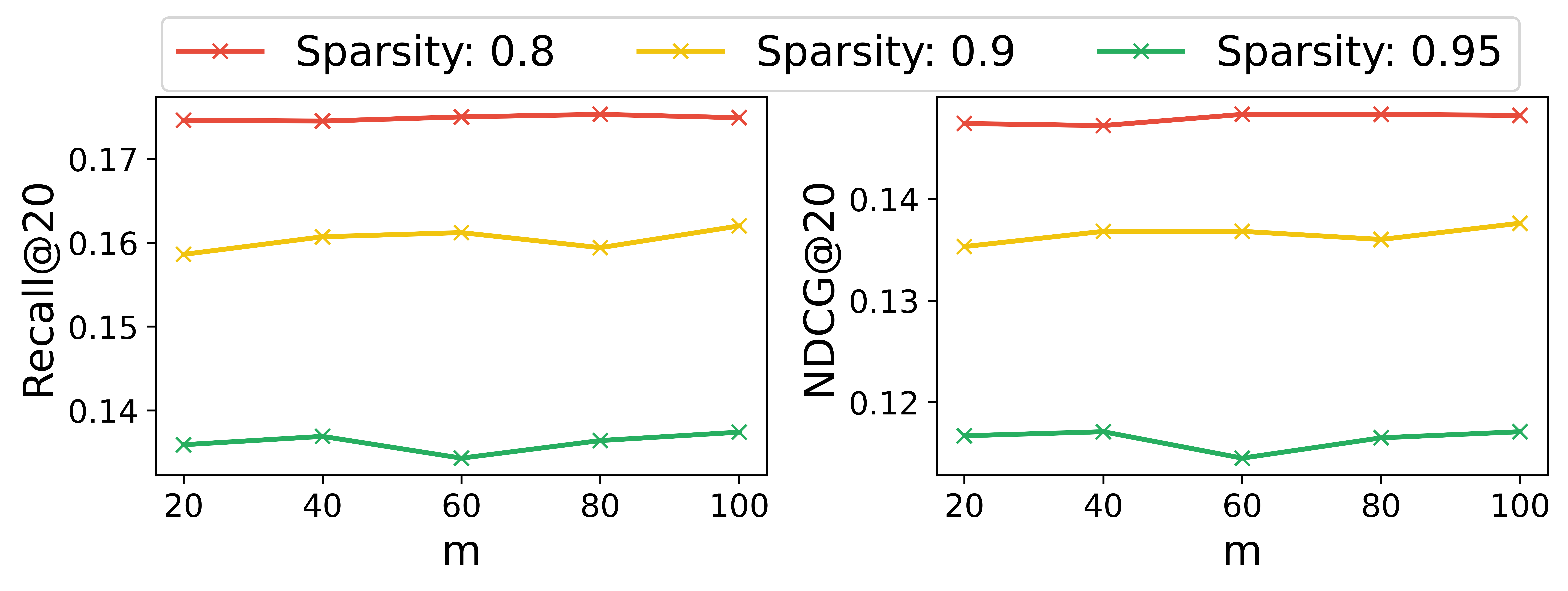} 
        \vspace{-2em}
        \caption{Effect of $m$ on Gowalla}
    \end{subfigure}
    \vspace{-1.3em}
    \caption{Sensitivity analysis of $m$ on Recall@20 and NDCG@20 with LightGCN as the backbone recommender.}
    \label{fig:m graph}
    \vspace{-1.3em}
\end{figure}

In each iteration, we sample $m$ candidate actions to choose from. We tune BET with $m \in \{20, 40, 60, 80, 100\}$ for both Yelp2018 and Gowalla. Then we show how the performance changes with different values of $m$ w.r.t. Recall@20 and NDCG@20 scores. As shown in Figure \ref{fig:m graph}, although BET is relatively insensitive to the value of $m$, better performance is achieved with $m = 100$ on both datasets.

\subsubsection{Number of Iterations $T$}

\begin{figure}[]
    \centering
    \begin{subfigure}{0.92\linewidth} 
        \centering
        \includegraphics[width=\linewidth]{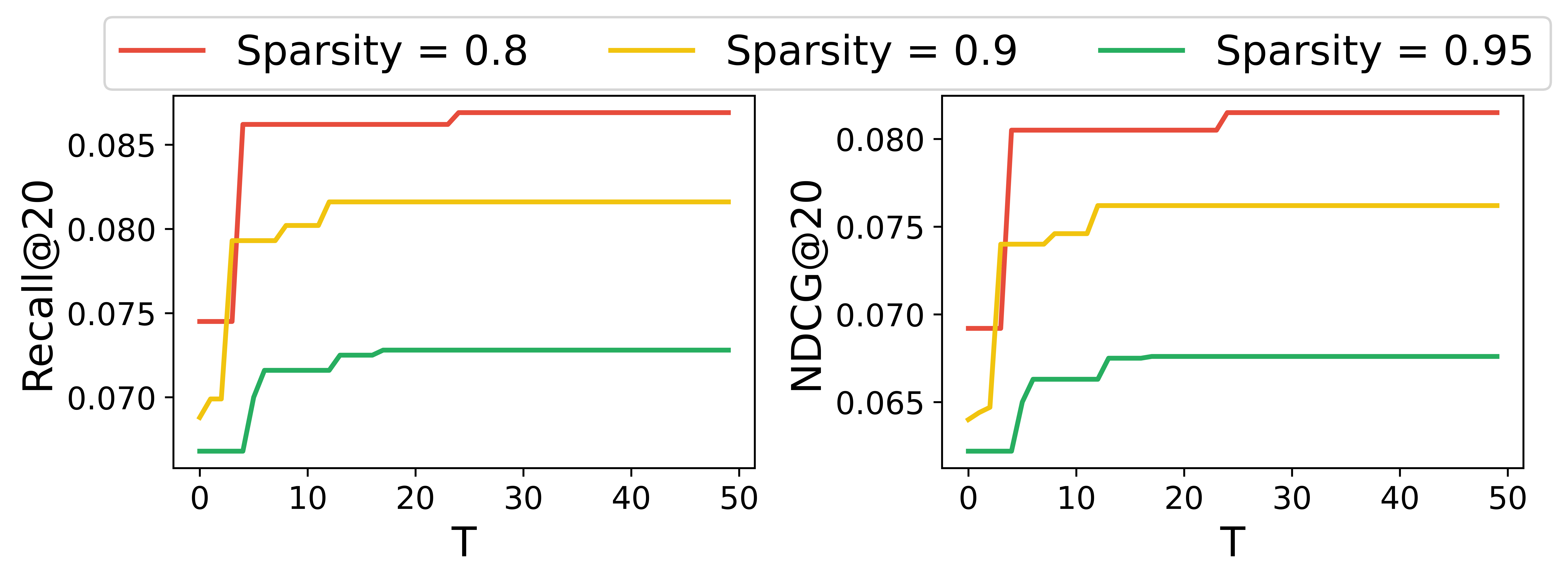}
        \vspace{-2em}
        \caption{Effect of $T$ on Yelp}
    \end{subfigure}
    \begin{subfigure}{0.92\linewidth} 
        \centering
        \includegraphics[width=\linewidth]{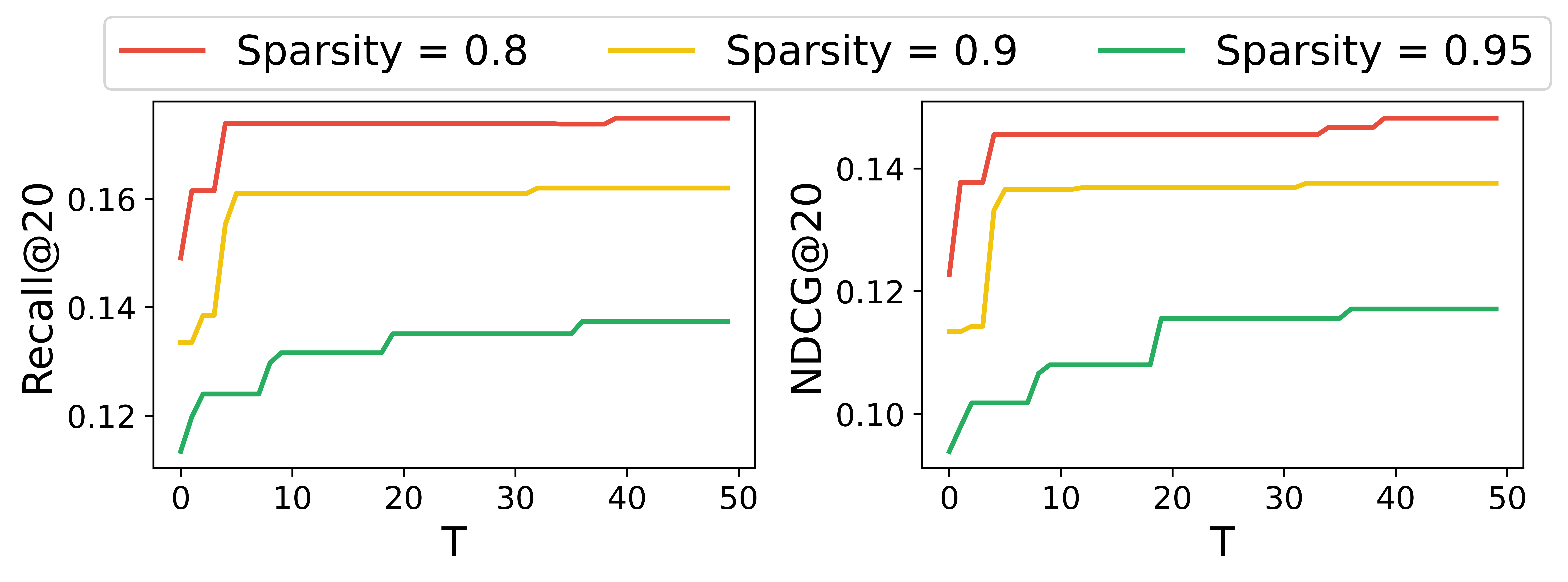} 
        \vspace{-2em}
        \caption{Effect of $T$ on Gowalla}
    \end{subfigure}
    \vspace{-1.3em}
    \caption{Sensitivity analysis of $T$ on Recall@20 and NDCG@20 with LightGCN as the backbone recommender.}
    \label{fig:T graph}
    \vspace{-1.3em}
\end{figure}

\begin{figure} 
    \centering
    \includegraphics[width=0.92\linewidth]{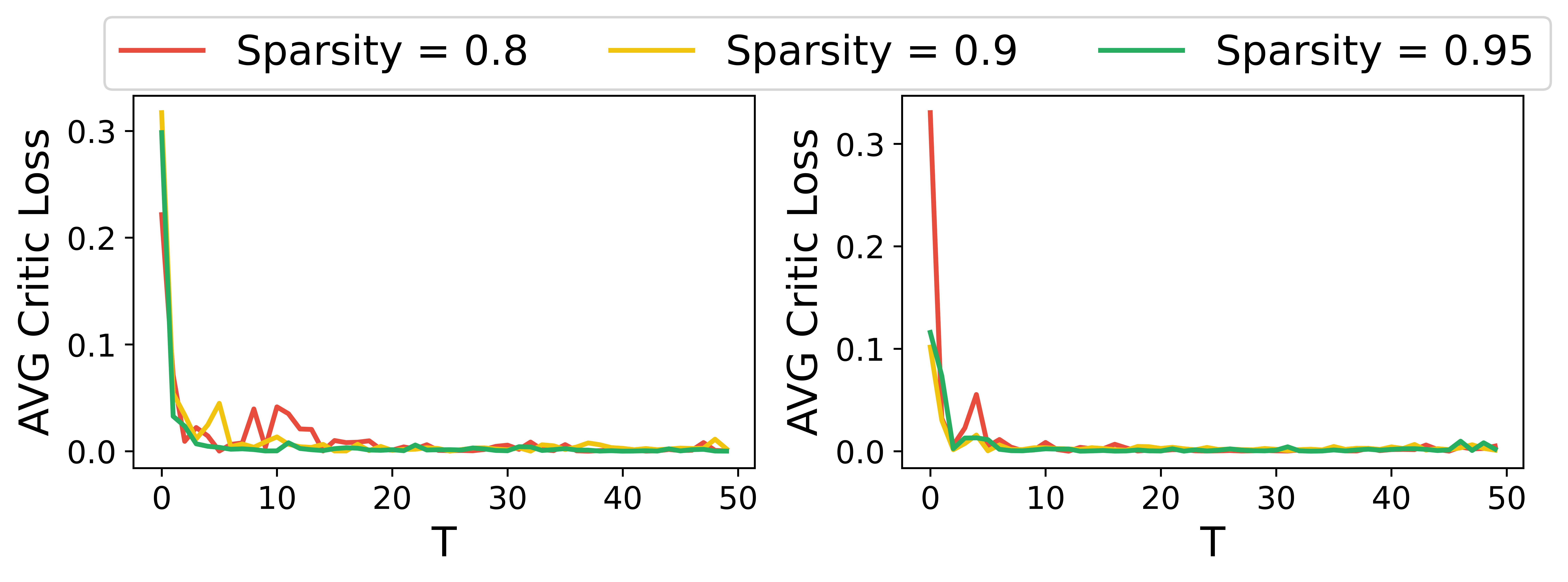}
    \vspace{-1.2em}
    \caption{Sensitivity analysis of $T$ on the fitness prediction loss with LightGCN as the backbone recommender on Yelp2018 (left) and Gowalla (right).} \label{fig:loss}
    \vspace{-1.5em}
\end{figure}

BET iteratively samples, selects and evaluates actions in search of the optimal one. One natural question that arises is how many iterations it needs to find the action that leads to peak recommendation performance. This question is crucial for understanding the effectiveness of BET. Therefore, we perform selective retraining as described in Section \ref{sec:retraining} on the resulting population $\mathcal{A}_t$ for each training iteration $t \in [0, T]$ and report the best recommendation performance in terms of Recall@20 and NDCG@20. We also report the average fitness prediction loss in each iteration. Figure \ref{fig:loss} shows that fitness predictor converges within the first 15 iterations. While the training loss of the fitness predictor is diminishing, BET witnesses a substantial improvement in its performance in the first 10-15 iterations, as shown in Figure \ref{fig:T graph}. Then performance continues to climb slowly after the 15th iteration until it reaches its peak around the 40th iteration and remains unchanged after that. In conclusion, setting $T = 50$ should suffice for BET to find the optimal action for all sparsity ratios.

\section{Conclusion and Future Works}
Latent factor recommenders typically employ vectorized embeddings with a uniform and fixed size, resulting in suboptimal performance and excessive memory complexity. To address this challenge, we have introduced BET, a model-agnostic algorithm to enable the selection of customized embedding sizes at the level of the embedding table for each user/item. This approach enhances the the embeddings while minimizing memory costs. Next, we will explore other importance signals such as model confidence \cite{qu2021human, qu2021human} or model weights \cite{Thimm1995Evaluating, Strom1997Sparse, Janowsky1989Pruning}.

\section*{Acknowledgment} This work is supported by the Australian Research Council under the streams of Future Fellowship (No. FT210100624), Discovery Project (No. DP190101985), and Discovery Early Career Research Award (No. DE200101465 and No. DE230101033).

% \nocite{*}
\normalem
\bibliographystyle{ACM-Reference-Format}
\bibliography{bet}

\end{document}